# COMPARISON OF RADIO PROPAGATION MODELS FOR LONG TERM EVOLUTION (LTE) NETWORK

Noman Shabbir[1], Muhammad T. Sadiq[2], Hasnain Kashif[3] and Rizwan Ullah[4]

[1]Department of Electrical Engineering, GC University,
Lahore, Pakistan
`noman.shabbir@gcu.edu.pk`
[2]Department of Electrical Engineering, Blekinge Institute of Tech.,
Karlskrona, Sweden
`mtsa09@bth.se`
[3]Department of Electrical Engineering, COMSATS Institute of IT,
Attock, Pakistan
`hasnain_kashif@comsats.edu.pk`
[4]Department of Electrical Engineering, COMSATS Institute of IT,
Attock, Pakistan
`rizwan_ullah@comsats.edu.pk`

## ABSTRACT

*This paper concerns about the radio propagation models used for the upcoming 4[th] Generation (4G) of cellular networks known as Long Term Evolution (LTE). The radio wave propagation model or path loss model plays a very significant role in planning of any wireless communication systems. In this paper, a comparison is made between different proposed radio propagation models that would be used for LTE, like Stanford University Interim (SUI) model, Okumura model, Hata COST 231 model, COST Walfisch-Ikegami & Ericsson 9999 model. The comparison is made using different terrains e.g. urban, suburban and rural area.SUI model shows the lowest path lost in all the terrains while COST 231 Hata model illustrates highest path loss in urban area and COST Walfisch-Ikegami model has highest path loss for suburban and rural environments.*

## KEYWORDS

*Long Term Evolution, LTE, Path loss, Propagation models.*

## 1. INTRODUCTION

Long Term Evolution (LTE) is the latest step in moving forward from the cellular 3[rd] Generation (3G) to 4[th] Generation (4G) services. LTE is often described as a 4G service but it is not fully compatible to 4G standards. An improved version of LTE known as LTE advance is a 4G compatible technology. Both LTE & LTE advance uses the same frequency band. LTE is based on standards developed by the 3rd Generation Partnership Project (3GPP). LTE may also be referred more formally as Evolved UMTS Terrestrial Radio Access (E-UTRA) and Evolved UMTS Terrestrial Radio Access Network (E-UTRAN). Even though 3GPP created the standards for family, the LTE standards are completely new with exceptions where it made sense.

The following are the main objectives for LTE [1].

- Increased downlink and uplink peak data rates

- Scalable bandwidth

- Improved spectral efficiency

- All IP network

- A standard's based interface that can support a multitude of user types

LTE will be having a downlink speed of 100 Mbps and an uplink of almost 50 Mbps. The channel will be having a scalable bandwidth from 1 MHz to 20 MHz while supporting both Frequency Division Duplex (FDD) and Time Division Duplex (TDD) [1]. These data rates can be further increased by employing multiple antennas both at the transmitter and receiver.

The selection of a suitable radio propagation model for LTE is of great importance. A radio propagation model describes the behavior of the signal while it is transmitted from the transmitter towards the receiver. It gives a relation between the distance of transmitter & receiver and the path loss. From this relation, one can get an idea about the allowed path loss and the maximum cell range. Path loss depends on the condition of environment (urban, rural, dense urban, suburban, open, forest, sea etc), operating frequency, atmospheric conditions, indoor/outdoor & the distance between the transmitter & receiver.

In this paper, a comparison is made between different radio propagation models in different terrains to find out the model having least path loss in a particular terrain and which has the highest.

## 2. RADIO PROPAGTION MODELS

### 2.1 SUI Model

Stanford University Interim (SUI) model is developed for IEEE 802.16 by Stanford University [2], [3]. It is used for frequencies above 1900 MHz. In this propagation model, three different types of terrains or areas are considered. These are called as terrain A, B and C. Terrain A represents an area with highest path loss, it can be a very dense populated region while terrain B represents an area with moderate path loss, a suburban environment. Terrain C has the least path loss which describes a rural or flat area. In Table 1, these different terrains and different factors used in SUI model are described.

Table 1: Different terrains & their parameters

| Parameters | Terrain A | Terrain B | Terrain C |
|:---:|:---:|:---:|:---:|
| a | 4.6 | 4 | 3.6 |
| b(1/m) | 0.0075 | 0.0065 | 0.005 |
| c(m) | 12.6 | 17.1 | 20 |

The path loss in SUI model can be described as

$$PL = A + 10\gamma \log\left(\frac{d}{d_o}\right) + X_f + X_h + S \qquad (1)$$

where $PL$ represents Path Loss in dBs, $d$ is the distance between the transmitter and receiver, $d_o$ is the reference distance (Here its value is 100), $X_f$ is the frequency correction factor, $X_h$ is the

Correction factor for BS height, $S$ is shadowing & $\gamma$ is the path loss component and it is described as

$$\gamma - a - bh_b + \frac{c}{h_b} \qquad (2)$$

where $h_b$ is the height of the base station and $a$, $b$ and $c$ represent the terrain for which the values are selected from the above table.

$$A = 20 \log \left( \frac{4\pi d_o}{\lambda} \right) \qquad (3)$$

where $A$ is free space path loss while $d_o$ is the distance between Tx and Rx and $\lambda$ is the wavelength. The correction factor for frequency & base station height are as follows:

$$X_f = 6 \log \left( \frac{f}{2000} \right) \qquad (4)$$

$$X_h = -10.8 \log \left( \frac{h_r}{2000} \right) \qquad (5)$$

where $f$ is the frequency in MHz, and $h_r$ is the height of the receiver antenna. This expression is used for terrain type A and B. For terrain C, the blow expression is used.

$$X_h = -20 \log \left( \frac{h_r}{2000} \right) \qquad (6)$$

$$S = 0.65(\log f)^2 - 1.3 \log (f) + \alpha \qquad (7)$$

Here, $\alpha$ = 5.2 dB for rural and suburban environments (Terrain A & B) and 6.6 dB for urban environment (Terrain C).

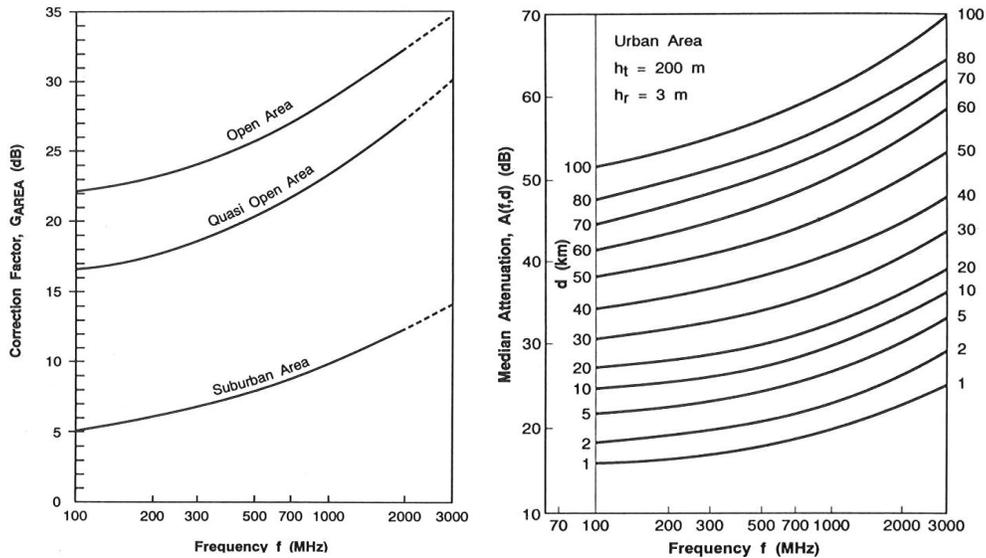

Figure 1: Median attenuation factor for Okumura Model

## 2.2 Okumura Model

Okumura model [4], [9] is one of the most commonly used models. Almost all the propagation models are enhanced form of Okumura model. It can be used for frequencies up to 3000 MHz. The distance between transmitter and receiver can be around 100 km while the receiver height can be 3 m to 10 m. The path loss in Okumura model can be calculated as

$$Pl(dB) = L_f + A_{m,n}(f,d) - G(h_t) - G(h_r) - G_{AREA} \quad (8)$$

Here $L_f$ is the free space path loss and it is calculated by the following expression:

$$L_f = -20 \log\left(\frac{\lambda}{4\pi d_o}\right) \quad (9)$$

While $G(h_t)$ and $G(h_r)$ are the BS antenna gain factor and receiver gain factors respectively. Their formulas are as follows:

$$G(h_b) = 20 \log\left(\frac{h_o}{200}\right) \quad (10)$$

$$G(h_r) = 10 \log\left(\frac{h_r}{3}\right) \quad (11)$$

where $h_b$ and $h_r$ are the heights of base station and receiver receptively. $A_{m,n}(f,d)$ is called as median attenuation factor. Different curves for median attenuation factor are used depending on the frequency and the distance between the transmitter and receiver. The area gain $G_{AREA}$ depends on the area being used and its graph along with median attenuation factor is depicted in Figure 1 [9].

## 2.3 Cost-231Hata Propagation Model

COST-231 Hata model is also known as COST Hata model. It is the extension of Hata model [6] and it can be used for the frequencies up to 2000 MHz. The expression for median path loss, PLU, in urban areas is given by

$$PL(dB) = 46.3 + 33.9 \log(f) - 13.82 \log(h_b) - a(h_r) + [44.9 - 6.55 \log(h_b)].$$
$$\log(d) + c \quad (12)$$

Here, $f$ represents the frequency in MHz, d denotes the distance between the transmitter & receiver, $h_b$ & $h_r$ the correction factors for base station height and receiver height respectively. The parameter $c$ is zero for suburban & rural environments while it has a value of 3 for urban area. The function $a(h_r)$ for urban area is defined as:

$$a(h_r) = 3.2(\log(11.75 h_r))^2 - 4.97 \quad (13)$$

and for rural & suburban areas its is as follows:

$$a(h_r) = (1.1 \log(f) - 0.7)h_r - (1.58f - 0.8) \quad (14)$$

## 2.4 COST-231 Walfisch-Ikegami Model

COST-231 Walfisch-Ikegami model is an extension of COST Hata model [6], [7]. It can be used for frequencies above 2000 MHz. When there is Line of Site (LOS) between the transmitter & receiver the path loss is given by the following formula:

$$PL = 42.64 + 26\log(d) + 20\log(f) \qquad (15)$$

While in Non-Line of Sight (NLOS) conditions, path loss is given as:

$$PL = L_o + L_{RTS} + L_{MSD} \qquad (16)$$

where $L_O$ is the attenuation in free-space and is described as:

$$Lo = 32.45 + 20\log(d) + 20\log(f) \qquad (17)$$

$L_{RTS}$ represents diffraction from rooftop to street, and is defined as:

$$L_{RTS} = -16.9 - 10\log(w) + 10\log(f) + 20\log(h_b - h_r) + L_{ori} \qquad (18)$$

Here $L_{ORI}$ is a function of the orientation of the antenna relative to the street $a$ (in degrees) and is defined as:

$$L_{ori} = \begin{cases} -10 + 0.354a & for\ 0 < a < 35 \\ 2.5 + 0.075(a-35) & for\ 35 < a < 55 \\ 4 - 0.114(a-55) & for\ 55 < a < 90 \end{cases} \qquad (19)$$

$L_{MSD}$ represents diffraction loss due to multiple obstacles and is specified as:

$$L_{MSD} = L_{BSH} + k_A + k_D\ \log(d) + k_F\log(f) - 9\log(s_b) \qquad (20)$$

Where

$$L_{BSH} = \begin{cases} -18\log(1 + h_t - h_b) & for\ h_t > h_b \\ 54 + 0.8(h_t - h_b)2d & for\ h_t \le h_b \\ & and\ d < 0.5km \end{cases} \qquad (21)$$

$$k_A = \begin{cases} 54 & for\ h_t > h_b \\ 54 + 0.8\left(h_t - h_b\right) & for\ h_t \le h_b \\ & and\ d > 0.5km \end{cases} \qquad (22)$$

$$k_D = \begin{cases} 18 + 15\left(\frac{h_t - h_b}{h_b}\right) & for \quad h_t > h_b \\ 18 & for \quad h_t \le h_b \\ & and\ d > 0.5\ km \end{cases} \qquad (23)$$

$$k_F = -4 + k\left(\frac{f}{924}\right) \qquad (24)$$

Here, $k$ = 0.7 for suburban centers and 1.5 for metropolitan centers.

## 2.5 Ericsson 9999 Model

This model is implemented by Ericsson as an extension of the Hata model [2], [11]. Hata model is used for frequencies up to 1900 MHz. In this model, we can adjust the parameters according to the given scenario. The path loss as evaluated by this model is described as:

$$PL = a_0 + a_1 \log(d) + a_2 \log(h_b) + a_3 \log(h_b) \log(d)$$
$$-3.2 (\log(11.75))^2 + g(f) \qquad (25)$$

where

$$g(f) = 44.49 \log(f) - 4.78 ((\log(f))^2 \qquad (26)$$

The values of $a_0$, $a_1$, $a_2$ and $a_3$ are constant but they can be changed according to the scenario (environment). The defaults values given by the Ericsson model are $a_0 = 36.2$, $a_1 = 30.2$, $a_2 = 12.0$ and $a_3 = 0.1$. The parameter $f$ represents the frequency.

## 3. SURVEY OF RELATED WORK

LTE is well positioned to meet the requirements of next-generation mobile networks for existing 3GPP operators. It will enable operators to offer high performance, mass market mobile broadband services, through a combination of high bit-rates and system throughput, in both the uplink and downlink and with low latency [1]. A comprehensive set of propagation measurements taken at 3.5 GHz in Cambridge, UK is used to validate the applicability of the three models mentioned previously for rural, suburban and urban environments. The results show that in general the SUI and the COST-231 Hata model over-predict the path loss in all environments. The ECC-33 model shows the best results, especially in urban environments [2]. They comparison of propagation models is also being done in [9] & [10].

## 4. PROBLEM STATEMENT AND MAIN CONTRIBUTION

Our research question is to find out the radio propagation model which will give us the least path loss in a particular terrain. The main problem is that LTE is using 1900 MHz and 2100 MHz frequency bands in different regions of the world. In some regions, frequencies of 700 MHz, 1800 MHz and 2600 MHz are also considered for LTE. For these frequency bands, many different radio propagation models are available that can be used in different terrains like urban, dense urban, suburban, rural etc. We will make a comparison between different radio propagation models and find out the model that is best suitable in a particular terrain. The comparison is made on the basis of path loss, antenna height and transmission frequency.

## 5. PROBLEM SOLUTION

In our simulation, two different operating frequencies 1900 MHz & 2100 MHz are used. The average building height is fixed to 15 m while the building to building distance is 50 m and street width is 25 m. All the remaining  parameters used in our simulations are described in Table 2.

Almost all the propagation models are available to be used both in LOS & NLOS environments. In our simulations, to make the scenario more practical, NLOS is used in urban, suburban & rural conditions. But LOS condition is being considered for rural area in COST 231 W-I model because it did not provide any specific parameters for rural area [10].

The empirical formulas of path loss calculation as described in the earlier section are used and the path loss is plotted against the distance for different frequencies & different BS heights. Figure 2 & Figure 3 shows the path loss for SUI model for 1900 MHz & 2100 MHz respectively. Similarly, Figure 4 & Figure 5 are for Okumura model for 1900 MHz & 2100 MHz respectively. In Figure 6, the path loss for COST 231 Hata model for 1900 MHz is shown. In Figure 7 & Figure 8, path loss for COST Walfisch-Ikegami Model is depicted for the same two frequencies.

Figure 9 & Figure 10 represents the path loss for Ericsson 9999 model.

Table 2: Simulation parameters

| Parameters | Values |
|---|---|
| Base station transmitter power | 43 dBm |
| Mobile transmitter power | 30 dBm |
| Transmitter antenna height | 30 m & 80 m in urban, suburban and rural area |
| Receiver antenna height | 3 m |
| Operating frequency | 1900 MHz & 2100 MHz |
| Distance between Tx & Rx | 5 km |
| Building to building distance | 50 m |
| Average building height | 15 m |
| Street width | 25 m |
| Street orientation angle | $30^0$ in urban and $40^0$ in suburban |
| Correction for shadowing | 8.2 dB in suburban and rural and 10.6 dB in urban area |

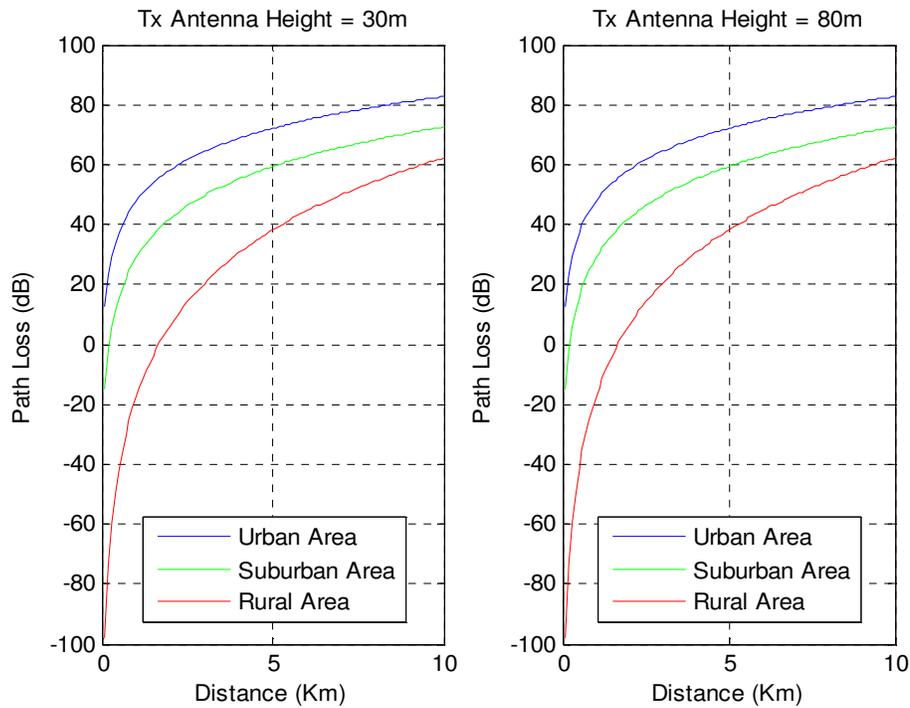

Figure 2: SUI model (for 1900 MHz)

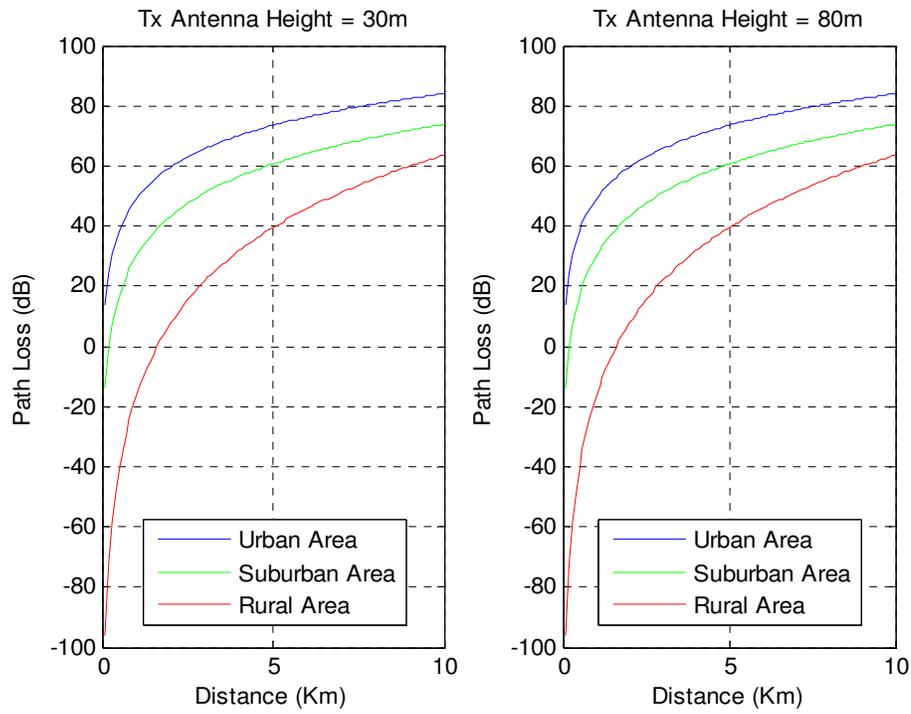

Figure 3: SUI model (for 2100 MHz)

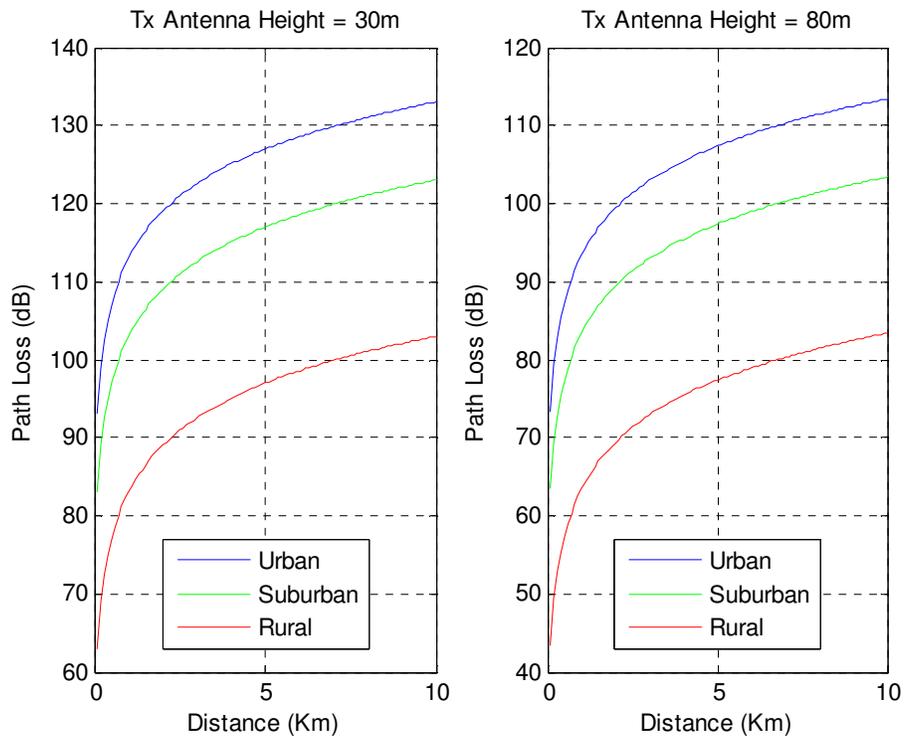

Figure 4: Okumura model (for 1900 MHz)

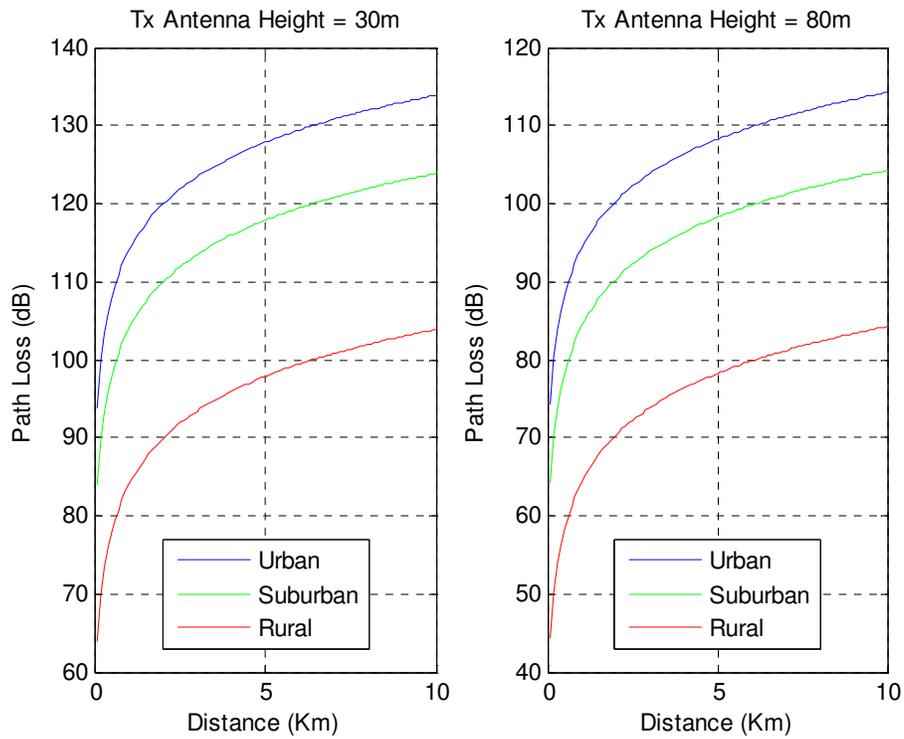

Figure 5: Okumura model (for 2100 MHz)

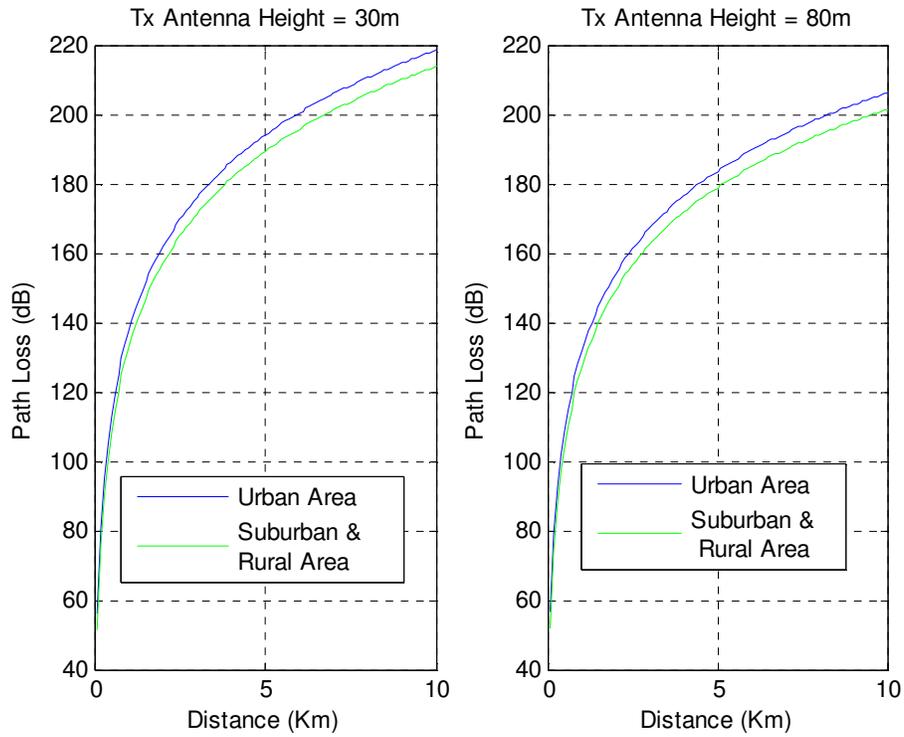

Figure 6: Hata COST 231 model (for 1900 MHz)

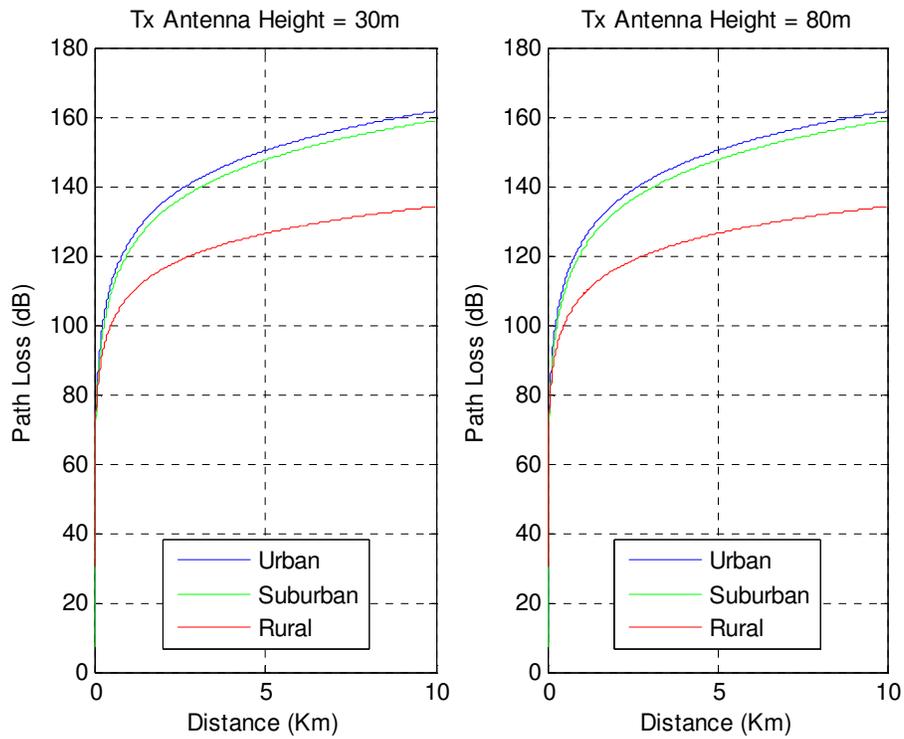

Figure 7: COST-231Walfisch-Ikegami model (for 1900 MHz)

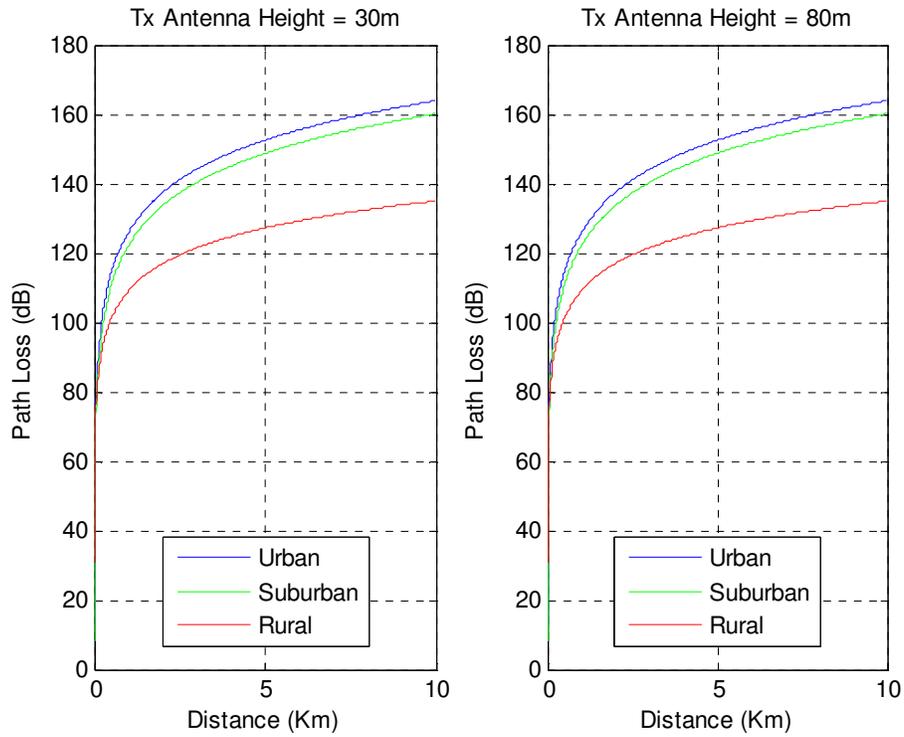

Figure 8: COST-231Walfisch-Ikegami model (for 2100 MHz)

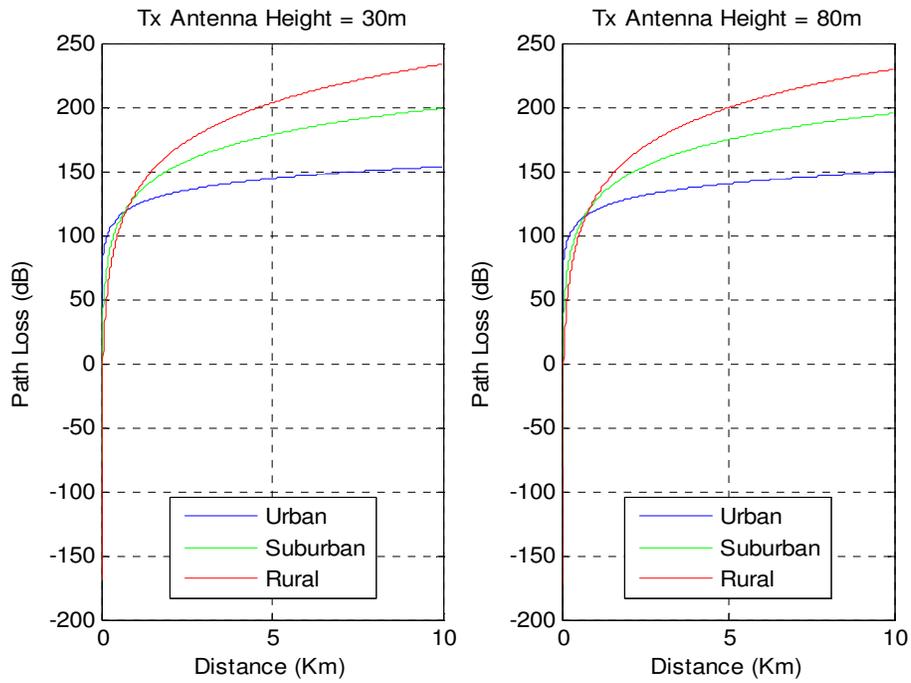

Figure 9: Ericsson 9999 model (for 1900 MHz)

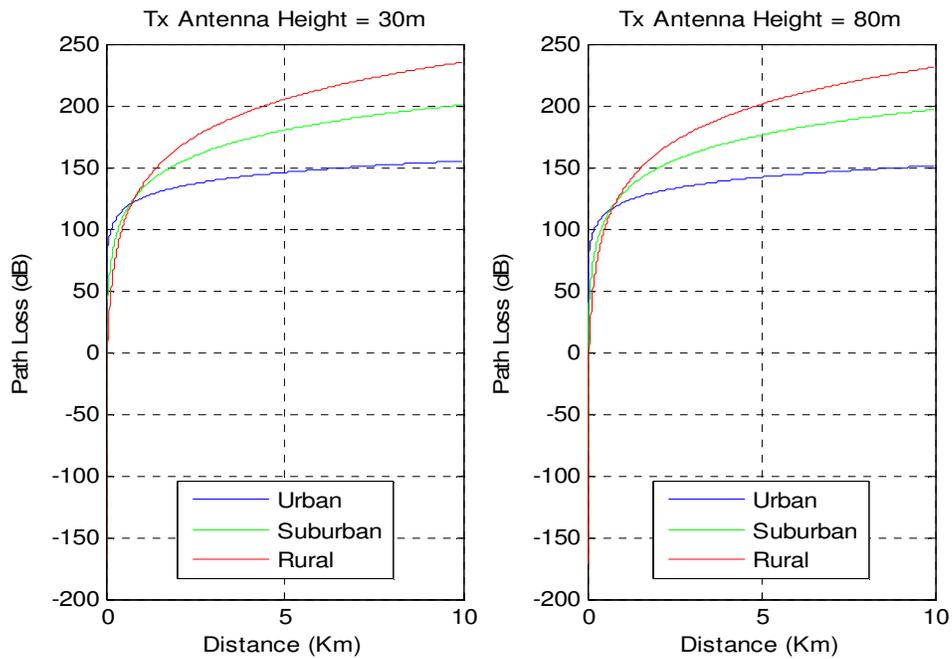

Figure 10: Ericsson 9999 model (for 2100 MHz)

## 6. CONCLUSION

The accumulated path losses for all the three urban, suburban and rural terrains are shown in Table 3. It can be seen from the table that SUI model has the lowest path loss prediction (72.17 dB to 73.43 dB) in urban environment for 1900 MHz & 2100 MHz respectively. While, COST 231 Hata model has the highest path loss (194.03 dB) for 1900 MHz in urban environment and Ericsson 999 model has the highest path loss of 145.83dB for 21000 MHz.

In suburban environment, the results are the same. SUI model shows the lowest path lost of 59.83 dB for 1900 MHz & 60.56 dB for 2100 MHz. COST 231 Hata model has the highest path lost of 189.32 dB for 1900 MHz & COST Walfisch-Ikegami model has a path loss of 148.69 dB for 2100 MHz.

In rural environment, SUI model has the lowest path lost of 38.20 dB for 1900 MHz & 39.46 dB for 2100 MHz. COST 231 Hata model has the highest path lost of 189.32 dB for 1900 MHz & COST Walfisch-Ikegami model has a path loss of 127.21 dB for 2100 MHz.

It can also be seen from the Table 3 that in suburban & rural environments, Ericsson 9999 model has more path loss than COST Walfisch-Ikegami model. Also, Ericsson 9999 model is having more path loss in suburban & rural environments than urban environment. That is something which is unrealistic. It is due to the fact that Ericsson 9999 model was mainly designed for urban and dense urban environments and it does not provide accurate information regarding suburban & rural areas. Hence, its values can be ignored. Another reason for using the COST Walfisch-Ikegami model is that it describes some additional parameters which are used to describe some environmental characteristics.

The BS height has no significant impact on the path loss of SUI model in all three terrains while all the other showed variation in path loss when their BS heights are changed. Okumura model has the highest variations.

Path losses for the frequencies of 700 MHz, 1800 MHz & 2600 MHz can be calculated using the above defined path loss equations.

Table 3: Comparison of propagation models

| Model | Frequency (MHz) | Distance (km) | BS Height (m) | Receiver Height (m) | Urban Path Loss (dB) | Suburban Path Loss (dB) | Rural Path Loss (dB) |
|---|---|---|---|---|---|---|---|
| SUI | 1900 | 5 | 30 | 3 | 72.17 | 59.83 | 38.20 |
| SUI | 1900 | 5 | 80 | 3 | 72.17 | 59.83 | 38.24 |
| SUI | 2100 | 5 | 30 | 3 | 73.43 | 60.56 | 39.46 |
| SUI | 2100 | 5 | 80 | 3 | 73.43 | 60.56 | 39.46 |
| Okumura | 1900 | 5 | 30 | 3 | 126.99 | 116.99 | 96.99 |
| Okumura | 1900 | 5 | 80 | 3 | 107.37 | 97.37 | 77.37 |
| Okumura | 2100 | 5 | 30 | 3 | 127.86 | 117.86 | 97.86 |
| Okumura | 2100 | 5 | 80 | 3 | 107.34 | 98.24 | 78.24 |
| Ericsson | 1900 | 5 | 30 | 3 | 144.31 | 178.38 | 203.26 |
| Ericsson | 1900 | 5 | 80 | 3 | 140.36 | 174.43 | 199.31 |
| Ericsson | 2100 | 5 | 30 | 3 | 145.83 | 179.90 | 204.79 |

| Ericsson | 2100 | 5 | 80 | 3 | 141.86 | 175.95 | 200.83 |
|---|---|---|---|---|---|---|---|
| COST 231 | 1900 | 5 | 30 | 3 | 194.03 | 189.32 | 189.32 |
| COST 231 | 1900 | 5 | 80 | 3 | 183.66 | 178.94 | 178.94 |
| Walfisch-Ikegami | 1900 | 5 | 30 | 3 | 150.20 | 147.51 | 126.35 |
| Walfisch-Ikegami | 1900 | 5 | 80 | 3 | 150.20 | 147.51 | 126.35 |
| Walfisch-Ikegami | 2100 | 5 | 30 | 3 | 152.47 | 148.69 | 127.21 |
| Walfisch-Ikegami | 2100 | 5 | 80 | 3 | 152.47 | 148.69 | 127.21 |

## Authors

**Noman Shabbir** was born in Lahore, Pakistan in 1985. He got his BS Computer Engineering degree from COMSATS Institute of Information Technology (CIIT), Lahore, Pakistan, in 2007 and MS Electrical Engineering (Radio Comm.) degree from Blekinge Institute of Technology, Sweden in 2009. His project on Unmanned Vehicles Arial Vehicles (UAV) got 3rd position in a National competition in 2006. He is currently working as a lecturer in the Dept. of Electrical Engineering., GC University, Lahore, Pakistan. His research interests are in the field of wireless networks and computer networks.

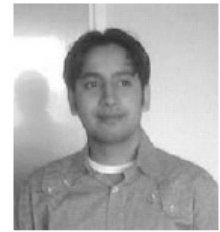

**Muhammad Tariq Sadiq** was born in Lahore, Pakistan, in 1985. He received Bachelor Degree in Electrical Engineering from COMSATS Institute of Information Technology (CIIT), Pakistan, in 2009. He is currently doing Masters Degree in Electrical Engineering with emphasis on telecommunication/ signal processing from Blekinge Institute of Technology (BTH), Karlskrona, Sweden. His research interests are in the field of cognitive radio, speech processing and wireless networks.

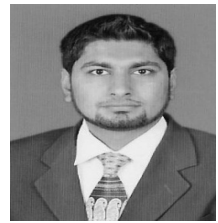

**Hasnain Kashif** was born in Tulamba, Distt. Khanewal, Pakistan. He has completed his BS in Computer Engineering from COMSATS Institute of Information Technology Lahore, Pakistan. He has done MSc Electrical Engineering (Radio communication) from BTH Karlskrona, Sweden. He has about two years experience in the field of telecommunication Companies in the department of Installation & Operation, Wireless Design and also RF Department. Now he is working as a Lecturer in Department of Electrical Engineering at COMSATS Institute of Information Technology Attock, Pakistan.

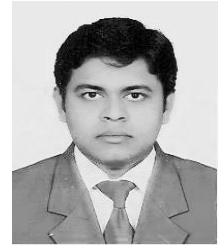

**Rizwan Ullah** was born in Oghi, Distt.  Mansehra, Pakistan. He has completed his BS in Electronic Engineering from COMSATS Institute of Information Technology Abbotabad, Pakistan. He is currently an MSc Electrical Engineering (Communication and Radar Technology) student at CIIT Attock, Pakistan. He has about 1.5 years experience as a Lab Engineer in Department of Electrical Engineering at COMSATS Institute of Information Technology Attock, Pakistan.

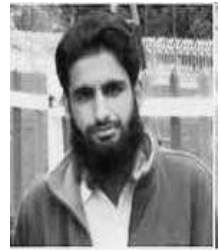

.